\documentclass[10pt,conference]{IEEEtran}
\IEEEoverridecommandlockouts
\usepackage{geometry}
\usepackage{cite}
\usepackage{amsmath,amssymb,amsfonts}
\usepackage{algorithmic}
\usepackage{algorithm} 
\usepackage[hidelinks,bookmarks=false]{hyperref}
\usepackage{graphicx}
\usepackage{textcomp}
\usepackage{xcolor}
\def\BibTeX{{\rm B\kern-.05em{\sc i\kern-.025em b}\kern-.08em
    T\kern-.1667em\lower.7ex\hbox{E}\kern-.125emX}}

\begin{document}

\title{Antenna Coding Optimization Based on \\Pixel Antennas for MIMO Wireless Power \\Transfer with DC Combining\\
}

\author{
	\IEEEauthorblockN{
		Yijun Chen\IEEEauthorrefmark{2}, 
		Shanpu Shen\IEEEauthorrefmark{3}\IEEEauthorrefmark{5}, 
		Tianrui Qiao\IEEEauthorrefmark{2}, 
            Hongyu Li\IEEEauthorrefmark{4},
            Jun Qian\IEEEauthorrefmark{2},
		and Ross Murch\IEEEauthorrefmark{2}} 
	\IEEEauthorblockA{\IEEEauthorrefmark{2}Department of Electronic and Computer Engineering, \\The Hong Kong University of Science and Technology, Hong Kong, China}
	\IEEEauthorblockA{\IEEEauthorrefmark{3}State Key Laboratory of Internet of Things for Smart City, University of Macau, Macau, China}	
    \IEEEauthorblockA{\IEEEauthorrefmark{5}Department of Electrical and Computer Engineering, University of Macau, Macau, China}	

     \IEEEauthorblockA{\IEEEauthorrefmark{4}Internet of Things Thrust, The Hong Kong University of Science and Technology (Guangzhou), Guangzhou, China\\
 E-mail: ychenoj@connect.ust.hk}
} 

\maketitle

\begin{abstract}
This paper investigates antenna coding based on pixel antennas as a new degree of freedom for enhancing multiple-input multiple-output (MIMO) wireless power transfer (WPT) systems. Antenna coding is closely related to the Fluid Antenna System (FAS) concept and further 
generalizes the radiation pattern reconfigurability. We first introduce a beamspace channel model to demonstrate reconfigurable radiation patterns enabled by antenna coders. By jointly optimizing the antenna coding and transmit beamforming with perfect channel state information (CSI), we exploit gains from antenna coding, transmit beamforming, and rectenna nonlinearity to maximize the output DC power. We adopt an alternating optimization approach with the quasi-Newton method and Successive Exhaustive Boolean Optimization (SEBO) method with warm-start to handle the transmit beamforming design and antenna coding design respectively. Finally, simulation results show that the proposed MIMO WPT system with pixel antennas achieves up to 15 dB gain in average output DC power compared with a conventional system with fixed antenna configuration, highlighting the potential of pixel antennas for boosting the WPT efficiency.  
\end{abstract}


\section{Introduction}
 Wireless power transfer (WPT) has emerged as a key technology for powering low-power devices in Internet of Things (IoT) and wireless sensor networks \cite{ClerckxFoundations}. With the massive explosion of low-power devices and reduction of power requirements, WPT offers a sustainable solution by delivering power without wires or battery replacements.  However, a critical challenge remains maximizing end-to-end power transfer efficiency,  i.e. maximizing the output direct current (DC) power at the energy receiver (ER) for a given transmit power at the energy transmitter (ET).

Prior works have focused on efficient rectenna design \cite{ullah2022review} and WPT signal design \cite{ClerckxFoundations} including waveform and beamforming designs to overcome this challenge. More recently, multiport rectenna systems with RF and DC combining schemes have been studied to enhance the output DC power by leveraging spatial diversity at the receiver \cite{shen2020beamforming}, \cite{shen2021joint}. However, most existing systems rely on antennas with fixed configuration, overlooking the potential of reconfigurable antennas in WPT systems.

To overcome this limitation, pixel antennas, as a highly reconfigurable antenna technology, provide a promising solution \cite{zhang2022highly}, \cite{zhang2022low} by controlling the RF switch states between the discretized sub-wavelength elements called  pixels. It enables dynamic reconfigurability of various antenna characteristics such as radiation pattern, frequency, and polarization \cite{song2014efficient},\cite{zheng2024design}. 
A closely-related research area to pixel antennas is the fluid antenna system (FAS) \cite{wong2020fluid}. By adjusting the antenna position within a linear region, FAS can improve channel conditions, leading to enhanced wireless communication performance. Pixel antennas have been used to implement FAS \cite{zhang2024pixel}, where the switch configuration is optimized to mimic the position adjustment. FAS has also found applications in WPT scenarios, such as simultaneous wireless information and power transfer (SWIPT) \cite{zhou2025fluid} and FAMA-assisted integrated data and energy transfer \cite{lin2025fluid}.

Nevertheless, the recently proposed concept of antenna coding empowered by pixel antennas \cite{shen2024antenna} further generalizes the  radiation pattern reconfigurability. Pixel antennas can be characterized by binary variables referred to as antenna coders, which represent the tunable switch states. Antenna coding has shown enhanced performance in SISO and MIMO communication systems \cite{shen2024antenna}. Therefore, it has great potential to address the limitation of previous WPT systems with fixed antenna configuration \cite{shen2017multiport}, \cite{shen2020directional}.

In this work, we apply the idea of antenna coding to MIMO WPT systems where pixel antennas are utilized at ET and ER with optimized antenna coders to boost power transfer efficiency. The contributions can be summarized as follows.

\textit{First}, we propose a MIMO WPT system model with pixel antennas based on the beamspace channel model and formulate the DC power maximization problem. \textit{Second}, we provide an alternating optimization approach to jointly optimize the antenna coding and transmit beamforming with DC combining scheme. \textit{Third}, we address the transmit beamforming design problem considering nonlinearity using an efficient quasi-Newton method. \textit{Fourth}, we evaluate the performance of the proposed antenna coding design for WPT, which significantly outperforms the conventional WPT system with fixed antenna configuration, with up to 15 dB gain. 

\textit{Organization}: Section II introduces the channel model with antenna coding. Section III presents the joint DC power maximization problem. Section IV develops an alternating optimization algorithm. Section V provides simulation results, and Section VI concludes the paper.

\textit{Notations}: Upper-case and lower-case bold letters represent matrices and vectors, respectively. A symbol without bold font denotes a scalar. ${\mathcal{E}}\{$·$\}$, $\mathbb{R}$ and $\mathbb{C}$ denote expectation, real and complex number sets, respectively. $[\mathbf{a}]_i$, $\mathbf{a}^*$ and $\|\mathbf{a}\|$ represent the $i$th element, conjugate, and the $l_{2}$-norm of a vector $\mathbf{a}$. $\mathbf{A}^{\mathrm{\mathit{T}}}$, $\mathbf{A}^{H}$, $[\mathbf{A}]_{i,:}$, and $[\mathbf{A}]_{i,j}$ denote the transpose, conjugate transpose, $i$th row, and $(i,j)$th element respectively. diag$ (a_{1},...,a_{N})$ denotes a diagonal matrix with entries $a_{1},...,a_{N}$. blkdiag$(\mathbf{a}_{1},...,\mathbf{a}_{N})$ denotes a block diagonal matrix formed by $\mathbf{a}_{1},...,\mathbf{a}_{N}$. $\mathbf{I}$ denotes the identity matrix.

\section{Antenna Coding with Pixel Antennas}
In this section, we briefly revisit the model of the pixel antenna with antenna coding, and introduce a MIMO beamspace channel model.

\subsection{Pixel Antenna and Antenna Coding}
\begin{figure}[t!]
    \centering
\includegraphics[width=\linewidth]{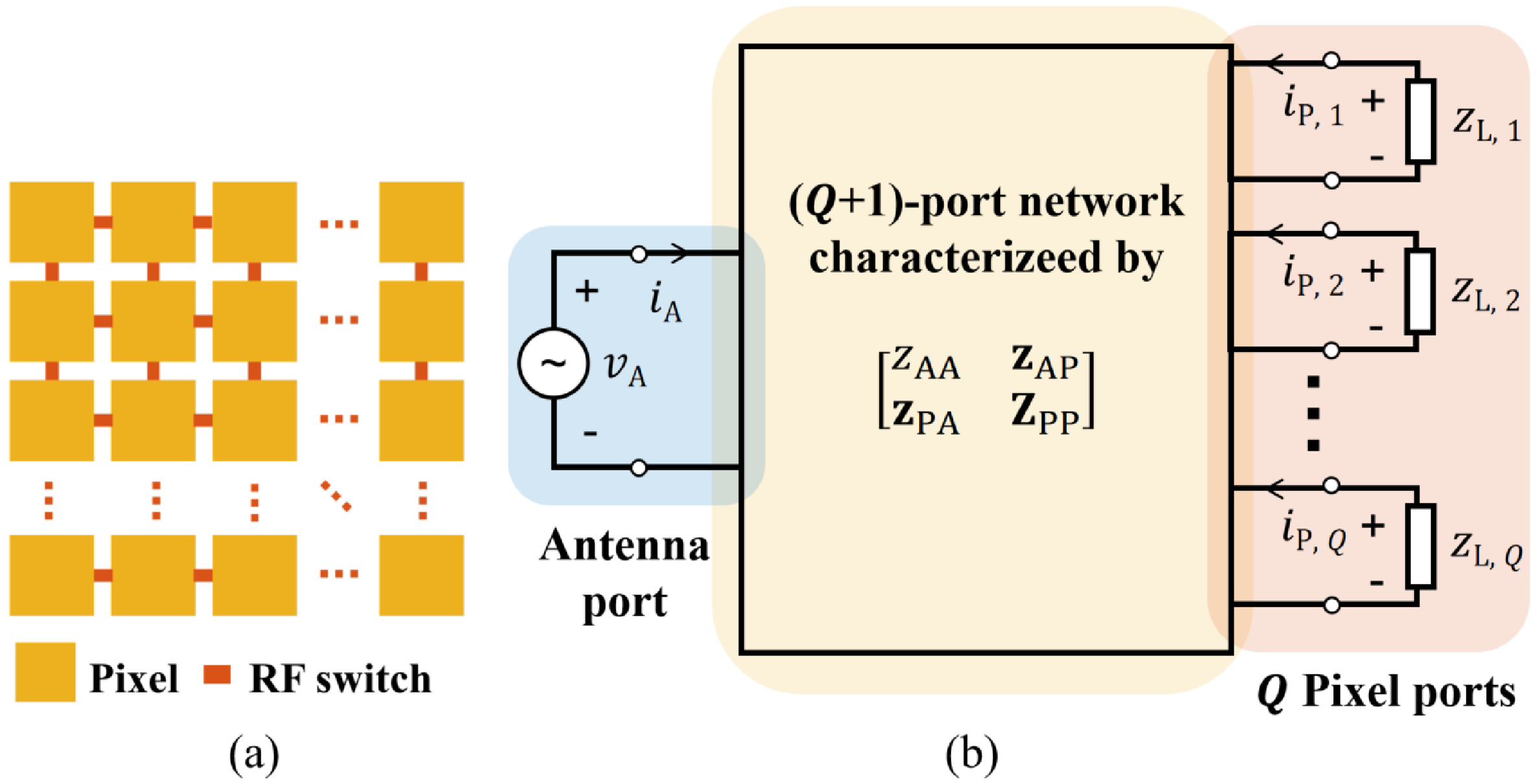} 
    \caption{Schematic of (a) a pixel antenna and (b) equivalent multiport network.}
    \label{pixel}
\end{figure}
As shown in Fig.~\ref{pixel}(a), the pixel antenna is based on a discretized radiation surface having $Q$ switches (with ON and OFF states) between small metallic sub-elements (marked in yellow) called pixels. According to the internal multiport method \cite{song2014efficient}, it can be regarded as a $(Q+1)$-port network with an antenna port and $Q$ pixel ports as illustrated in Fig. 1(b). The $(Q+1)$-port network can be characterized by its impedance matrix $\mathbf{Z}\in\mathbb{C}^{(Q+1)\times(Q+1)}$, which is written as
\begin{equation}
\mathbf{Z=}\begin{bmatrix}z_{\mathrm{AA}} & \mathbf{z}_{\mathrm{AP}}\\
\mathbf{z}_{\mathrm{PA}} & \boldsymbol{\mathbf{Z}}_{\mathrm{PP}}
\end{bmatrix},
\end{equation}
where $z_{\mathrm{AA}}\in\mathbb{C}$ and $\mathbf{Z}_{\mathrm{PP}}\in\mathbb{C}^{Q\times Q}$ represent the self impedance for the antenna port and pixel ports respectively, and $\mathbf{z}_{\mathrm{AP}}=\mathbf{z}_{\mathrm{PA}}^T\in\mathbb{C}^{Q\times1}$ are the trans-impedance between the antenna port and pixel ports. 

The states of switches between pixels are represented by a binary vector $\mathrm{\mathbf{b}=}\mathrm{[\mathit{\mathrm{\mathit{b_{1},...,b_{\mathit{Q}}}}}]^{\mathit{T}}\in\mathbb{C}^{\mathit{Q\times1}}}$ referred to as an antenna coder. We model the switches by a load impedance matrix $\mathbf{Z}_{\mathrm{L}}(\boldsymbol{\mathrm{b}})=\mathrm{diag}(\mathrm{\mathit{z}_{L,1},...,\mathit{z}_{L,\mathit{Q}}})\in\mathbb{C}^{\mathit{Q\times Q}}$, where each load impedance $\mathit{z}_{\mathrm{L},q}$ can be open or short-circuit depending on the switch states as
\begin{equation}
\mathit{z}_{\mathrm{L},q}=\begin{cases}
\infty, & \mathrm{if}\:b_{q}=1,\\
0, & \mathrm{if}\:b_{q}=0.
\end{cases}\label{eq}
\end{equation}

When exciting the antenna port with current $i_{\mathrm{A}}\in\mathbb{C}$, the currents through pixel ports $\mathbf{i}_{\mathrm{P}}(\mathbf{b})=[i_{\mathrm{P},1},...,i_{\mathrm{P},Q}]\in\mathbb{C}^{\mathit{Q\times1}}$ can be found as
\begin{equation}
\mathbf{i}_{\mathrm{P}}(\mathbf{b})=-(\mathbf{Z}_{\mathrm{PP}}+\mathbf{Z}_{\mathrm{L}}(\mathbf{b}))^{-1}\mathbf{z}_{\mathrm{PA}}i_{\mathrm{A}},
\end{equation}
which is coded by the antenna coder $\mathbf{b}$.

Accordingly, the radiation pattern $\mathbf{e}(\mathbf{b})\in\mathbb{C}^{\mathit{\mathrm{2}K\times\mathrm{1}}}$ of the pixel antenna is given by
\begin{equation}
\mathrm{\mathbf{e}(\mathbf{b})=\mathbf{\mathbf{e}}_{\mathrm{A}}\mathit{i}_{\mathrm{A}}+\sum_{\mathit{q}=1}^{\mathit{Q}}\mathbf{\mathbf{e}}_{\mathrm{\mathrm{P,\mathit{q}}}}\mathit{i}_{P,\mathit{q}}(\mathbf{b})=\mathbf{E}_{\mathrm{oc}}\mathbf{i}(\mathbf{b})},
\end{equation}
where $\mathbf{e}_{\mathrm{A}}\in\mathbb{C}^{\mathit{\mathrm{2\mathit{K}\times1}}}$ and $\mathbf{\mathbf{e}}_{\mathrm{\mathrm{P,\mathit{q}}}}\in\mathbb{C}^{2K\times 1}$ are the open-circuit radiation patterns of the antenna port and the $q$th pixel port under unit-current excitation with $\theta$ and $\phi $ polarization components over $K$ spatial angles respectively, $\mathbf{E_{\mathrm{oc}}=\mathrm{[\mathbf{e}_{\mathrm{A}},\mathbf{e}_{\mathrm{\mathrm{P},1}},...,\mathbf{e}_{\mathrm{\mathrm{P},\mathit{Q}}}]}}\in\mathbb{C}^{\mathrm{\mathit{\mathrm{2}K\times(Q+\mathrm{1})}}}$ collects the open-circuit radiation patterns of all ports, and $\mathbf{i}(\mathbf{b})=[i_{\mathrm{A}};\mathbf{i}_{\mathrm{P}}(\mathbf{b})]\in\mathbb{C}^{(Q+1)\times1}$ is the current vector for all ports. By optimizing the antenna coder $\mathbf{b}$ among $2^Q$ switch configurations, the radiation characteristics of the pixel antenna can be dynamically adjusted to adapt to the channel.

\subsection{Beamspace Channel Model}

\begin{figure}[t!]
    \centering
    \includegraphics[width=\linewidth]{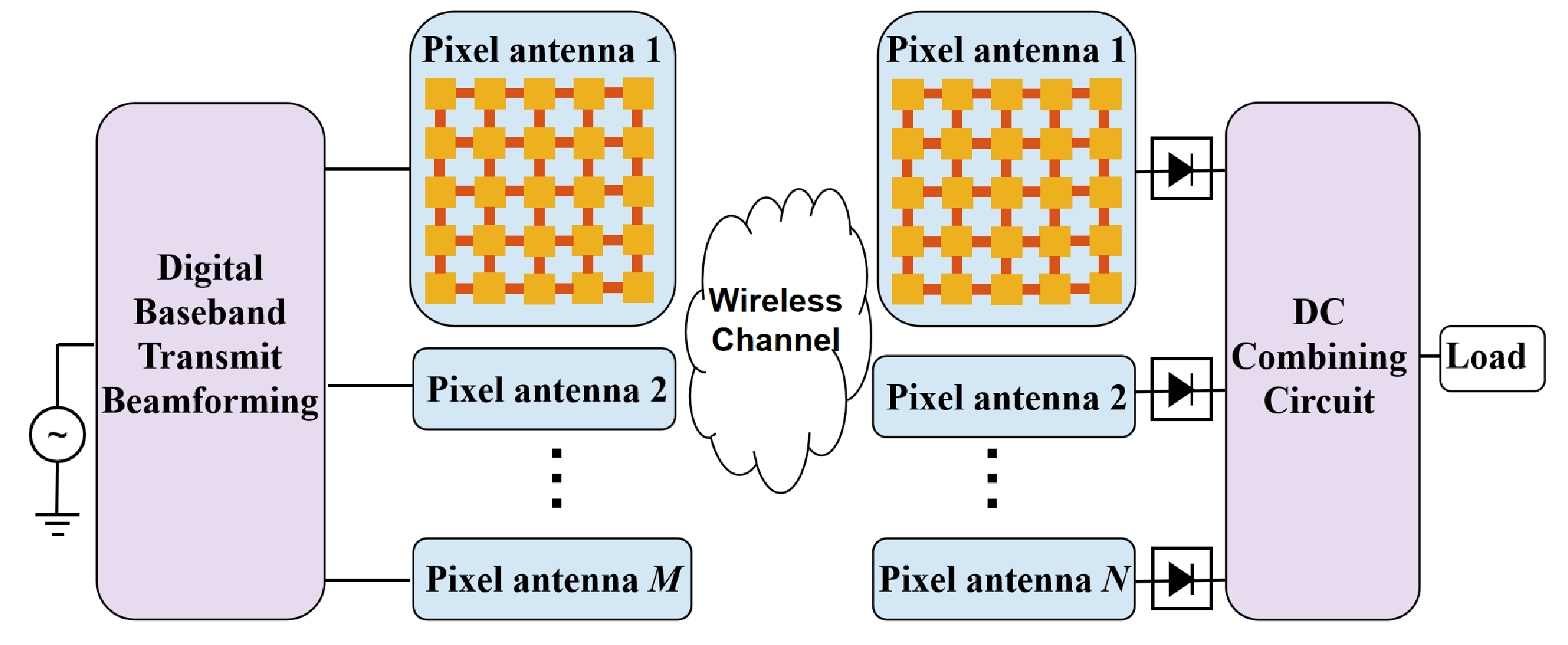} 
    \caption{Diagram of a MIMO WPT system with $M$ transmit and $N$ receive pixel antennas deployed at the energy transmitter and receiver under DC combining scheme.}
    \label{2}
\end{figure}

We consider a MIMO WPT system with $M$ transmit pixel antennas at the ET and $N$ receive pixel antennas at the ER as shown in Fig.~\ref{2}. The antenna coders of each transmit and receive pixel antenna are denoted by $\mathbf{b}_{\mathrm{T,\mathit{m}}},m=1,2,...,M$ and $\mathbf{b}_{\mathbf{\mathrm{R,\mathit{n}}}},n=1,2,...,N$, which are collected into matrices $\mathbf{B}_{\mathrm{T}}=\left[\mathbf{b}_{\mathrm{T},1},\ldots,\mathbf{b}_{\mathrm{T},M}\right]\in\mathbb{C}^{Q\times M}$ and $\mathbf{B}_{\mathrm{R}}=\left[\mathbf{b}_{\mathrm{R},1},\ldots,\mathbf{b}_{\mathrm{R},N}\right]\in\mathbb{C}^{Q\times N}$. Correspondingly, we denote the radiation patterns of the transmit and receive pixel antennas respectively as $\mathbf{E_{\mathrm{T}}\mathrm{(}\mathbf{B}_{\mathrm{T}}\mathrm{)}}=[\mathbf{e}_{\mathrm{T,1}}(\mathbf{b}_{\mathrm{T},1}),...,\mathbf{e}_{\mathrm{T,\mathit{M}}}(\mathbf{b}_{\mathrm{T},M})]\in\mathbb{C}^{\mathrm{\mathit{\mathrm{2}K\times M}}}$ and $\mathbf{E_{\mathrm{R}}\mathrm{(\mathbf{B}_{\mathrm{R}})}}=[\mathbf{e}_{\mathrm{R,1}}(\mathbf{b}_{R,1}),...,\mathbf{e}_{\mathrm{R,\mathit{N}}}(\mathbf{b}_{\mathrm{R},N})]\in\mathbb{C}^{\mathrm{\mathit{\mathrm{2}K\times N}}}$ with each column being normalized as $||\mathbf{e}_{\mathrm{T,\mathit{m}}}(\mathbf{b}_{\mathrm{T},m})||=1,\forall m$ and $||\mathbf{e}_{\mathrm{R,\mathit{n}}}(\mathbf{b}_{\mathrm{R},n})||=1,\forall n$. Hence, the MIMO beamspace channel model $\mathbf{H}\left(\mathbf{B}_{\mathrm{T}},\mathbf{B}_{\mathrm{R}}\right)\in\mathbb{C}^{N\times M}$ of the pixel antenna system can be given as
\begin{equation}
\mathbf{H}\left(\mathbf{B}_{\mathrm{T}},\mathbf{B}_{\mathrm{R}}\right)=\mathbf{E}_{\mathrm{R}}^{T}\left(\mathbf{B}_{\mathrm{R}}\right)\mathbf{H}_{\mathrm{V}}\mathbf{E}_{\mathrm{T}}\left(\mathbf{B}_{\mathrm{T}}\right),
\end{equation}
where  $\mathbf{H}_{\mathrm{V}}\in\mathbb{C}^{2\mathit{K}\times2\mathit{K}}$ is the virtual channel matrix collecting entries of channel gains between all pairs of angle of departure (AoD) and angle of arrival (AoA) among the $K$ spatial angles considered. Denoting $\mathbf{i}_{\mathrm{T},m}(\mathbf{b}_{\mathrm{T},m})\in\mathbb{C}^{(Q+1)\times 1}$ and $\mathbf{i}_{\mathrm{R},n}(\mathbf{b}_{\mathrm{R},n})\in\mathbb{C}^{(Q+1)\times 1}$ as the current vectors at the $m$th transmit and $n$th receive pixel antenna, the columns in $\mathbf{E}_{\mathrm{T}}\left(\mathbf{B}_{\mathrm{T}}\right)$ and  $\mathbf{E}_{\mathrm{R}}\left(\mathbf{B}_{\mathrm{R}}\right)$ are respectively given as 
\begin{align}
\mathbf{e}_{\mathrm{T},m}(\mathbf{b}_{\mathrm{T},m})&=\mathrm{\mathbf{E}_{\mathbf{\mathrm{oc}}}\mathbf{i}_{\mathrm{T},m}(\mathbf{b}_{\mathrm{T},m})},\\\mathbf{e}_{\mathrm{R},n}(\mathbf{b}_{\mathrm{R},n})&=\mathbf{E}_{\mathbf{\mathrm{oc}}}\mathbf{i}_{\mathrm{R},n}(\mathbf{b}_{\mathrm{R},n}).
\end{align}

To demonstrate the extra degrees of freedom introduced by pixel antennas, we  perform singular value decomposition (SVD) for $\mathrm{\mathbf{E}_{oc}=\mathbf{USV}^{\mathit{H}}}$, where $ N_{\mathrm{eff}}=\mathrm{rank}(\mathbf{E}_{\mathrm{oc}})$ is regarded as the effective aerial degrees of freedom (EADoF) in the beamspace domain, indicating the number of orthogonal radiation patterns a pixel antenna can provide, $\mathrm{\mathbf{U}}\in\mathbb{C}^{\mathrm{\mathit{\mathrm{2}K\times N_{\mathrm{eff}}}}}$ and $\mathbf{V}\in\mathbb{C}^{\mathrm{\mathit{(Q\mathrm{+1})\times N_{\mathrm{eff}}}}}$ are semi-unitary matrices with $\mathbf{U^{\mathrm{\mathit{H}}}U=V^{\mathrm{\mathit{H}}}V=I}$, and $\mathbf{S}\in\mathbb{R}^{N_{\mathrm{eff}}\times N_{\mathrm{eff}}}$ is a diagonal matrix containing $N_{\mathrm{eff}}$ non-zero singular values. Consequently, the orthogonal radiation patterns of each pixel antenna can be represented by the $N_{\mathrm{eff}}$ columns of $\mathrm{\mathbf{U}}$, which enables synthesizing of various radiation patterns. With the extra degrees of freedom, the MIMO system with pixel antennas can be equivalently regarded as a conventional MIMO system with $\mathit{N_{\mathrm{t}}=MN_{\mathrm{eff}}}$ and $\mathit{N_{\mathrm{r}}=NN_{\mathrm{eff}}}$ spatially separated antennas. 

Assuming a 2-D uniform power angular spectrum (PAS), the AoDs and AoAs vary only with the azimuth angles $\phi_{\mathrm{t}}^{k}$ and $\phi_{\mathrm{r}}^{k}$ respectively. We then denote the unitary orthogonal basis matrices across multiple transmit and receive pixel antennas as $\mathbf{E_{\mathrm{bs,T}}}=[\mathbf{U_{\mathrm{T,1}},...,U_{\mathrm{T,\mathit{M}}}}]\in\mathbb{C}^{\mathrm{\mathit{\mathrm{2}K\times N_{\mathrm{t}}}}}$ and $\mathbf{E_{\mathrm{bs,R}}}=[\mathbf{U_{\mathrm{R,1}},...,U_{\mathrm{R,\mathit{N}}}}]\in\mathbb{C}^{\mathrm{\mathit{\mathrm{2}K\times N_{\mathrm{r}}}}}$ respectively. The orthogonal basis of the $m$th transmit pixel antenna $\mathbf{U}_{\mathrm{T,\mathit{m}}}\in\mathbb{C}^{\mathrm{\mathit{\mathrm{2}K\times N_{\mathrm{eff}}}}}$ follows 
\begin{equation}
 [\mathbf{U}_{\mathrm{T,\mathit{m}}}]_{k',:}=[\boldsymbol{\mathbf{U}}]_{k',:}[\Phi_{m}]_{k'}, \forall k'=1,...,2K, 
\end{equation}
 where $\Phi_{m}=[\Delta\phi_{m}^{1};...;\Delta\phi_{m}^{K};\Delta\phi_{m}^{1};...;\Delta\phi_{m}^{K}]\in\mathbb{C}^{\mathrm{\mathit{\mathrm{2}K\times\mathrm{1}}}}$ and $\Delta\phi_{m}^{k}$ is given by $\Delta\phi_{m}^{k}=\mathrm{exp}(j2\pi(m-1)d_{\mathrm{t}}\mathrm{sin}\phi_{\mathrm{t}}^{k}/\lambda),\forall k$ with wavelength $\lambda$ and physical spacing $\mathit{d_{\mathrm{t}}}$ between pixel antennas. Similarly, the orthogonal basis of the $n$th receive pixel antenna $\mathbf{U}_{\mathrm{R,\mathit{n}}}\in\mathbb{C}^{\mathrm{\mathit{\mathrm{2}K\times N_{\mathrm{eff}}}}}$ can be defined.

Therefore, we can define the pattern coders for the $m$th transmit and $n$th receive pixel antenna respectively as
\begin{align}
\mathbf{w}_{\mathrm{T},m}(\mathbf{b}_{\mathrm{T},m}) &
= \mathbf{S} \mathbf{V}^\mathit{H} \mathbf{i}_{\mathrm{T},m}(\mathbf{b}_{\mathrm{T},m}) \label{eq:pt}, \\
\mathbf{w}_{\mathrm{R},n}(\mathbf{b}_{\mathrm{R},n}) &= \mathbf{S} \mathbf{V}^\mathit{T} \mathbf{i}^{*}_{\mathrm{R},n}(\mathbf{b}_{\mathrm{R},n}), \label{eq:pr}
\end{align}
where the power is constrained by $\mathrm{\left\Vert \mathrm{\mathbf{w}_{\mathrm{T,\mathit{m}}}(\mathbf{b}_{\mathrm{T,\mathit{m}}}\mathrm{)}}\right\Vert =1}$ and $\left\Vert \mathbf{w}_{\mathrm{R,\mathit{n}}}(\mathbf{b}_{\mathrm{R,\mathit{n}}}\mathrm{)}\right\Vert =1$. Instead of weighting over different antennas, the pattern coding enables highly reconfigurable patterns by forming the beams across the $N_{\mathrm{eff}}$ orthogonal basis patterns. The pattern coder $\mathbf{w}_{\mathrm{T},m}(\mathbf{b}_{\mathrm{T},m}) = \mathbf{U}_{\mathrm{T},m}^\mathit{H} \mathbf{e}_{\mathrm{T},m}(\mathbf{b}_{\mathrm{T},m})$ can also be regarded as a projection of the pattern $\mathbf{e}_{\mathrm{T,\mathit{m}}}(\mathbf{b}_{\mathrm{T,\mathit{m}}})$ onto the orthogonal basis matrix $\mathbf{U}_{\mathrm{T},m}$. With no mutual coupling between the adjacent pixel antennas, the pattern coders across all transmit and receive pixel antennas $\mathbf{W_{\mathrm{T}}\mathrm{(}B_{\mathrm{T}}\mathrm{)}}\in\mathbb{C}^{\mathrm{\mathit{N_{\mathrm{t}}\times M}}}$ and $\mathbf{W_{\mathrm{R}}\mathrm{(}B_{\mathrm{R}}\mathrm{)}}\in\mathbb{C}^{\mathrm{\mathit{N_{\mathrm{r}}\times N}}}$ can be represented as block diagonal matrices respectively as
\begin{align}
\mathbf{W_{\mathrm{T}}\mathrm{(}B_{\mathrm{T}}\mathrm{)}}&=\mathrm{blkdiag}(\mathbf{w_{\mathrm{T,1}}\mathrm{(}b_{\mathrm{T,1}}\mathrm{)}},...,\mathbf{w_{\mathrm{T,\mathit{M}}}\mathrm{(}b_{\mathrm{T,\mathit{M}}}\mathrm{)}}),\\
\mathbf{W_{\mathrm{R}}\mathrm{(}B_{\mathrm{R}}\mathrm{)}}&=\mathrm{blkdiag}(\mathbf{w_{\mathrm{R,1}}\mathrm{(}b_{\mathrm{R,1}}\mathrm{)}},...,\mathbf{w_{\mathrm{R,\mathit{N}}}\mathrm{(}b_{\mathrm{R,\mathit{N}}}\mathrm{)}}).
\end{align}

Hence, substituting the original expression (5) with (11) and (12), the overall beamspace channel matrix can be equivalently rewritten as
\begin{equation}
\mathbf{H}(\mathbf{B}_{\mathrm{T}},\mathbf{B}_{\mathrm{R}})=\mathbf{W_{\mathit{\mathrm{R}}}^{\mathit{\mathrm{\mathit{H}}}}(B_{\mathrm{R}})H_{\mathrm{C}}W_{\mathrm{T}}(B_{\mathrm{T}})}\label{eq},
\end{equation}
where $\mathbf{H}_{\mathrm{C}}\in\mathbb{C}^{\mathrm{\mathit{N_{\mathrm{r}}\times N_{\mathrm{t}}}}}$ can be regarded as a multi-antenna channel matrix that indicates the channel gain between the $\mathrm{\mathit{N_{\mathrm{t}}}}$ and $\mathrm{\mathit{N_{\mathrm{r}}}}$ equivalent spatially separated antennas and can be written as 
\begin{equation}
\mathbf{H_{\mathrm{C}}}=\mathbf{\mathbf{E}_{\mathrm{bs,R}}^{\mathit{T}}H_{\mathrm{V}}\mathbf{E_{\mathrm{bs,T}}}}\label{eq}.
\end{equation}

Compared with the virtual channel representation, the multi-antenna channel matrix $\mathbf{H_{\mathrm{C}}}$ has a compact size, which is beneficial for reducing both the channel estimation and optimization overheads. We consider a rich scattering propagation environment with Rayleigh fading so that $[\mathbf{H}_{\mathrm{C}}]_{i,j}\,\forall i,j$ can be modeled as independent and identically distributed (i.i.d.) random variables following the complex Gaussian distribution $\mathcal{CN}(0,1)$.

\section{Pixel Antenna Empowered MIMO WPT System}
In this section, the nonlinear rectenna model is first briefly introduced. Based on this, we propose the MIMO WPT system model with DC combining scheme and formulate the DC power maximization problem with perfect channel state information (CSI). 

\subsection{Nonlinear Rectenna Model}
\begin{figure}[t!]
    \centering
    \includegraphics[width=0.8\linewidth]{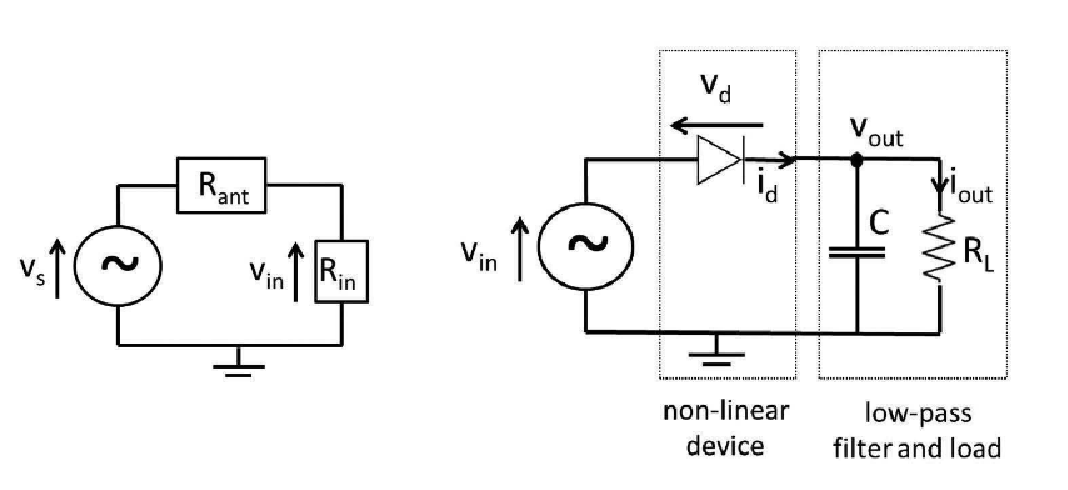} 
    \caption{Antenna equivalent circuit (left) and a single series diode rectifier (right).}
    \label{fig3}
\end{figure}
Consider a rectenna model as shown in Fig.~\ref{fig3}, each receive pixel antenna is assumed to be lossless and can be modeled as an equivalent voltage source $\mathrm{\mathit{v}_{s}(\mathit{t})}$ in series with antenna impedance $\mathit{R}_{\mathrm{ant}}$. Given the received signal $\mathit{y}(\mathit{t})$, the average received power at the antenna is denoted  as $\mathrm{\mathit{P}_{av}={\mathcal{E}}\{\mathit{y}(\mathit{t})^{2}\}}$. In the low-power region (e.g. below mW input power), the nonlinear model using Taylor expansion at zero quiescent point to the $n_{0}$th-order term can be conveniently utilized. The final output DC voltage of the rectifier is then approximated with
\begin{equation}
\mathrm{\mathit{v}_{out}=\sum_{\mathit{i}\,even,\mathit{i}\geq2}^{\mathit{n_{\mathrm{0}}}}\beta_{\mathit{i}}{\mathcal{E}}\{\mathit{y}(\mathit{t})^{\mathit{i}}\}}\label{eq},
\end{equation}
where $\beta_{\mathit{i}}=\frac{\mathit{R}_{\mathrm{ant}}^{\mathit{i}/2}}{\mathit{i}!(\mathit{I}_{\mathrm{d}}\mathit{v}_{\mathrm{t}})^{\mathit{i}-1}}$, $\mathrm{\mathit{v}_{t}}$ is the thermal voltage, $\mathit{I}_{\mathrm{d}}$ denotes the ideality factor, and $\mathrm{\mathit{n_{\mathrm{0}}}=4}$ is verified as a tractable tradeoff choice \cite{shen2020beamforming}.

\subsection{MIMO WPT System Model}
We denote the digital baseband transmit beamformer as $\mathbf{p}=[p_{1},...,p_{M}]^{T}\in\mathbb{C}^{M\times1}$ with power constraint $\frac{1}{2}||\mathbf{p}||^{2}\leq P$, where $P$ is the transmit power, and $\mathrm{\mathit{p_{m}},\forall}m$ is the complex weight at the $m$th transmit pixel antenna. The transmitted signal vector $\mathbf{x}(t)\in\mathbb{R}^{M\times1}$ after conventional digital baseband beamforming is given as
\begin{equation}
\mathrm{\mathbf{x}(\mathit{t})=\Re\{\mathbf{p}e^{\mathit{j\omega_{c}t}}\}}\label{eq},
\end{equation}
where $\omega_{c}$ denotes the  center frequency. The received signal $\mathbf{y}(\mathit{t})=[y_{1}(t),...,y_{N}(t)]\in\mathrm{\mathbb{R}}^{\mathrm{\mathit{N}\times1}}$ is then expressed as
\begin{align}
\mathbf{y}(\mathit{t})&=\Re\{\mathbf{H}\left(\mathbf{B}_{\mathrm{T}},\mathbf{B}_{\mathrm{R}}\right)\mathbf{p}e^{j\omega_{c}t}\}\\
&=\Re\{\mathbf{W_{\mathit{\mathrm{R}}}^{\mathit{H}}(B_{\mathrm{R}})H_{\mathrm{C}}W_{\mathrm{T}}(B_{\mathrm{T}})}\mathbf{p}e^{j\omega_{c}t}\}
\label{eq:yn}.
\end{align}

Using the nonlinear rectenna model, the output DC voltage of the $n$th rectenna is
\begin{equation}
v_{\mathrm{out},n}	\mathrm{=\sum_{\mathit{i\,\mathrm{even},\,i\geq\mathrm{2}}}^{\mathit{n_{\mathrm{0}}}}\beta_{\mathit{i}}{\mathcal{E}}\left\{ \mathit{y_{n}(t)^{\mathit{i}}}\right\} }\label{eq},
\end{equation}
where we have $\mathrm{{\mathcal{E}}\left\{ \mathit{y_{n}(t)^{\mathit{i}}}\right\} }=\zeta_{i}\left|[\mathbf{H}\left(\mathbf{B}_{\mathrm{T}},\mathbf{B}_{\mathrm{R}}\right)\mathbf{p}]_{n}\right|^{i}$, $\zeta_{\mathit{i}}=1/{2\pi}\int_{0}^{2\pi}\sin^{\mathit{i}}\mathit{t}\:\textrm{d}\mathit{t}$ and specifically $\zeta_{2}=\frac{1}{2}$, and $\zeta_{4}=\frac{3}{8}$.

With DC combining scheme as illustrated in Fig.~\ref{2}, the received RF signals at each pixel antenna are individually rectified by $N$ rectifiers and combined with a DC combining circuit such as a MIMO switching DC-DC converter to give the total output DC power $\mathrm{\mathit{P}_{\mathrm{out}}^{\mathrm{DC}}=\sum_{\mathit{n}=1}^{\mathit{N}}\mathit{v_{\mathrm{out},n}^{\mathrm{2}}}/\mathit{R_{\mathrm{L}}}}$. We aim to jointly optimize the antenna coders and transmit beamformers with perfect CSI, which yields the DC power maximization problem
\begin{align}
\underset{\mathbf{p},\, \mathbf{B}_{\mathrm{T}},\,\mathbf{B}_{\mathrm{R}}}{\max} 
& \sum_{n=1}^{N} \frac{v_{\mathrm{out},n}^2}{R_{\mathrm{L}}} \label{pro}\\
\mathrm{s.t.} \quad
& \frac{1}{2} \left\| \mathbf{p} \right\|^2 \leq P, \\
& [\mathbf{B}_{\mathrm{T}}]_{i,j} \in \{0,1\}, \,\forall i,j, \\
& [\mathbf{B}_{\mathrm{R}}]_{i,j} \in \{0,1\}, \, \forall i,j.\label{con}
\end{align}

This is a non-convex problem with highly-coupled optimization variables. The required CSI can be accurately acquired by channel estimation using beamspace pilot symbols and feedbacks strategy.

\section{Antenna Coding and Beamforming Design}
To deal with problem \eqref{pro}-\eqref{con}, we adopt the alternating optimization approach in this section and propose algorithms for the decoupled sub-problems to optimize antenna coder and transmit beamformer alternatively.

\subsection{Transmit Beamforming Optimization}
 We first consider fixed antenna coders $\mathrm{\mathbf{B}_{T}^{\star}}$ and $\mathrm{\mathbf{B}_{\mathrm{R}}^{\star}}$ to yield fixed channel $\mathbf{H}\left(\mathbf{B}_{\mathrm{T}}^{\star},\mathbf{B}_{\mathrm{R}}^{\star}\right)$, so that the sub-problem is written
\begin{align}
\underset{\mathbf{p}}{\max} 
& \quad \sum_{n=1}^{N} \frac{v_{\mathrm{out},n}^2}{R_{\mathrm{L}}} \label{quasi}\\
\mathrm{s.t.} 
& \quad \frac{1}{2} \left\| \mathbf{p} \right\|^2 \leq P\label{power}.
\end{align}

It is straightforward to show that it gives the same optimal solution with constraint $\frac{1}{2} \left\| \mathbf{p} \right\|^2 = P$. Then, we represent $\mathbf{p}=\sqrt{2P}\frac{\mathbf{p}_{0}}{||\mathbf{p}_{0}||}$ with auxiliary variable $\mathbf{p}_{0}$. Therefore, the transmit beamforming design problem can be transformed into an unconstrained optimization problem for $\mathbf{p}_{0}$, where the quasi-Newton method can be applied to find a stationary point solution. The optimized transmit beamformer adapts to the channel conditions considering nonlinearities. Performing SVD of the channel $\mathrm{\mathbf{H}=\mathbf{U}_{h}\mathbf{\Sigma}_{h}\mathbf{V}_{h}^{\mathit{H}}}$, the initialization point can be found as $\mathbf{p}_{\mathrm{init}}=\mathrm{\mathbf{v}_{1}\sqrt{2\mathit{P}}}$, where $\boldsymbol{\mathrm{v}}_{1}$ is the right singular vector corresponding to the largest singular value $\sigma_{1}$ in $\mathbf{\Sigma}_{\mathrm{h}}$.

\subsection{Antenna Coding Optimization}
Given optimized transmit beamformer $\mathrm{\mathbf{p}^{\star}}$ for now, we then consider the antenna coder design problem
\begin{align}
\underset{\mathbf{B}_{\mathrm{T}},\,\mathbf{B}_{\mathrm{R}}}{\mathrm{max}}\;\;\;
& \sum_{\mathit{n}=1}^{\mathit{N}}\frac{\mathit{v}_{\mathrm{out},\mathit{n}}^{2}}{R_{\mathrm{L}}}  \label{26}\\
\mathrm{s.t.}\quad
& [\mathbf{B}_{\mathrm{T}}]_{i,j} \in \{0,1\}, \, \forall i,j, \label{27}\\
& [\mathbf{B}_{\mathrm{R}}]_{i,j} \in \{0,1\}, \, \forall i,j.\label{28}
\end{align}

This is an NP-hard binary optimization problem, which can be solved by heuristic searching algorithms such as successive exhaustive Boolean optimization (SEBO) \cite{shen2016successive}. The SEBO optimization process is divided into two stages: 1) cyclic exhaustive searching of each block (blocksize $J$) of the antenna coders until convergence for an iteration, and 2) randomly flipping bits in the solution to look for potential better local optimal points. The complexity of SEBO can be given as $\mathbb{\mathcal{O}}(\mathrm{\mathit{I_{\mathrm{e}}}}\times2^{\mathrm{\mathit{J}}})$, where $I_{\mathrm{e}}$ is the number of iterations. To enable more effective search for problem \eqref{26}-\eqref{28} which has many local-optimums and highly depends on the initial point selection, we utilize the warm-start initialization strategy to run SEBO for multiple times using the obtained best solution as the initial point of the next-round optimization.  In this way, the  transmit beamformer and antenna coders can be alternatively optimized while fixing the other until convergence. The proposed alternating optimization algorithm is summarized in Algorithm 1.

\begin{algorithm}[t!]
    \caption{Alternating Optimization for Transmit Beamforming and Antenna Coding Design}
    \label{alg:ao_algorithm}
    \begin{algorithmic}[1]
        \STATE \textbf{Initialize:} $i = 0$, initial transmit beamformer $\mathbf{p}_{\mathrm{init}}$, initial antenna coders $\mathbf{B}_{\mathrm{T}}^{(0)}$, $\mathbf{B}_{\mathrm{R}}^{(0)}$, convergence threshold $\epsilon$, max iterations $i_{\max}$;
        \REPEAT
            \STATE $i \gets i + 1$;
            \STATE \textbf{Update Channel:} Compute the beamspace channel $\mathbf{H}(\mathbf{B}_{\mathrm{T}}^{(i-1)},\mathbf{B}_{\mathrm{R}}^{(i-1)})=\mathbf{W_{\mathit{\mathrm{R}}}^{\mathit{\mathrm{\mathit{H}}}}}(\mathbf{B}_{\mathrm{R}}^{(i-1)})\mathbf{H}_{\mathrm{C}}\mathbf{W}_{\mathrm{T}}(\mathbf{B}_{\mathrm{T}}^{(i-1)})$;
            \STATE \textbf{Optimize Transmit Beamformer $\mathbf{p}$:}
            \STATE \quad Fix $\mathbf{B}_{\mathrm{T}}^{(i-1)}$, $\mathbf{B}_{\mathrm{R}}^{(i-1)}$;
            \STATE \quad Use quasi-Newton to solve \eqref{quasi} and \eqref{power};
            \STATE \quad Obtain $\mathbf{p}^{(i)}$ via normalization: 
                    $\mathbf{p}^{(i)} =\sqrt{2P}\frac{\mathbf{p}_{0}}{||\mathbf{p}_{0}||}$;
            \STATE \textbf{Optimize Antenna Coder $( \mathbf{B}_{\mathrm{T}}, \mathbf{B}_{\mathrm{R}})$:}
            \STATE \quad Fix $\mathbf{p}^{(i)}$;
            \STATE \quad Run SEBO with warm-start to solve \eqref{26}-\eqref{28};
            \STATE \quad Obtain $\mathbf{B}_{\mathrm{T}}^{(i)}$, $\mathbf{B}_{\mathrm{R}}^{(i)}$;
        \UNTIL{objective change $< \epsilon$ or $i = i_{\max}$}
        \STATE \textbf{Output:} Optimal $\mathbf{p}^{\star} = \mathbf{p}^{(i)}$, $\mathbf{B}_{\mathrm{T}}^{\star} = \mathbf{B}_{\mathrm{T}}^{(i)}$, $\mathbf{B}_{\mathrm{R}}^{\star} = \mathbf{B}_{\mathrm{R}}^{(i)}$.
    \end{algorithmic}
\end{algorithm}

\section{Performance Evaluation}
In this section, we evaluate the performance of the proposed MIMO WPT system using pixel antennas.

\subsection{Simulation Setup and Pixel Antenna}
\begin{figure}[t!]
  \centering
  \includegraphics[width=\linewidth]{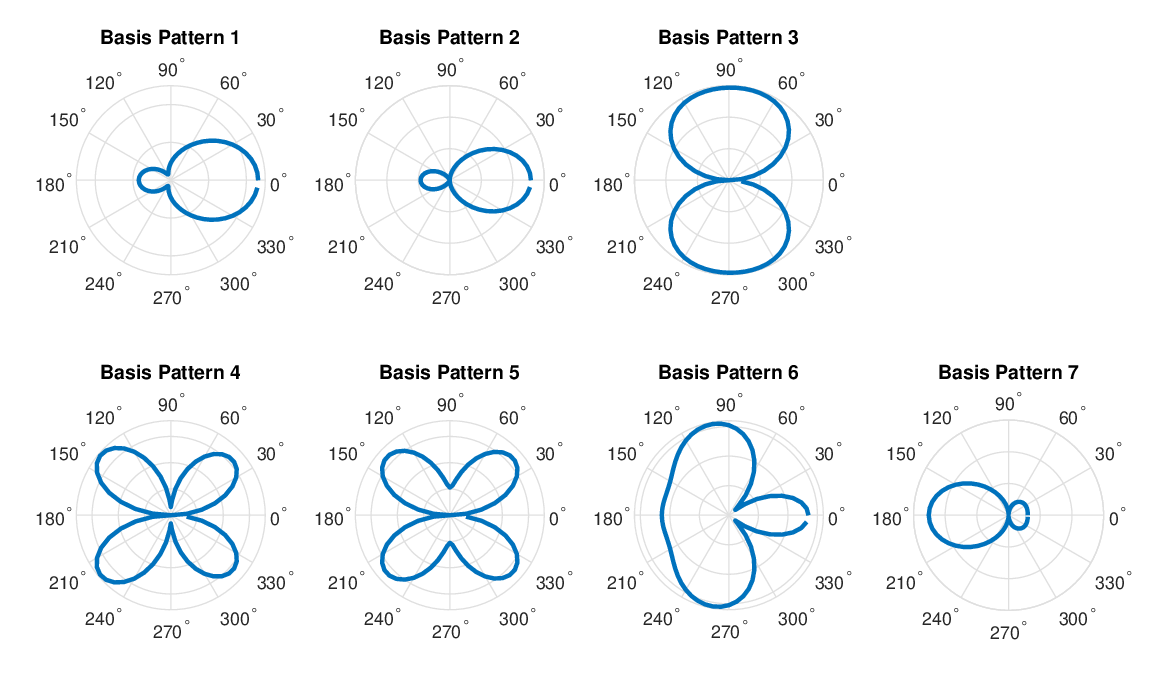}
  \caption{The orthogonal basis patterns provided by the pixel antenna in \cite{shen2024antenna}.}
  \label{pattern}
\end{figure}

We consider a rich scattering environment with a 2-D uniform PAS and $K=72$ sampled angles.  Identical and linearly spaced pixel antennas are deployed at both ER and ET.  Similar to \cite{shen2024antenna}, we simulate a pixel antenna sized $0.5\lambda\times0.5\lambda$ with one antenna port and $Q=39$ RF switches, where  $\lambda=125\,\mathrm{mm}$ at 2.4 GHz operation frequency. The impedance matrix $\mathbf{Z}\in\mathbb{C}^{(Q+1)\times(Q+1)}$ and the open-circuit radiation pattern matrix $\mathbf{E}_{\mathrm{oc}}\in\mathbb{C}^{2K\times(Q+1)}$ can be efficiently obtained using full-wave electromagnetic solver CST Microwave Studio through a single simulation. The rank of $\mathbf{E}_{\mathrm{oc}}$ is determined by the number of singular values that cumulatively account for over 99.8\% of the total power. This yields $N_{\mathrm{eff}}=7$ orthogonal patterns provided by the pixel antenna as shown in Fig.~\ref{pattern}. 

For the rectenna model, the parameters are given as $\mathrm{\mathit{R}_{\mathrm{ant}}=50\,\Omega}$, $\mathit{R}_{\mathrm{L}}=5\,\mathrm{k}\Omega $, $\mathit{I}_{\mathrm{d}}=1.05$, and $\mathrm{\mathit{v}_{t}}=25\,\mathrm{mV}$. We assume a multipath fading environment with a 36 dBm transmit power and 66 dB path loss, which yields an average received power of -30 dBm. We assume that entries of $\mathbf{H}_{\mathrm{C}}$ are modeled as i.i.d. complex Guassian random variables. Monte Carlo method with 1000 channel realizations is used to find the average output DC power. 

We evaluate the performance of the proposed MIMO WPT system with antenna coding based on pixel antennas with Algorithm 1, which is referred to as adaptively optimized (OPT) beamforming. Our benchmark is the transmit beamforming scheme toward the largest singular value direction of the channel, which is denoted as SVD beamforming optimal for the average received RF power.  

\subsection{System Performance}

\begin{figure}[t!]
    \centering   \includegraphics[width=0.8\linewidth]{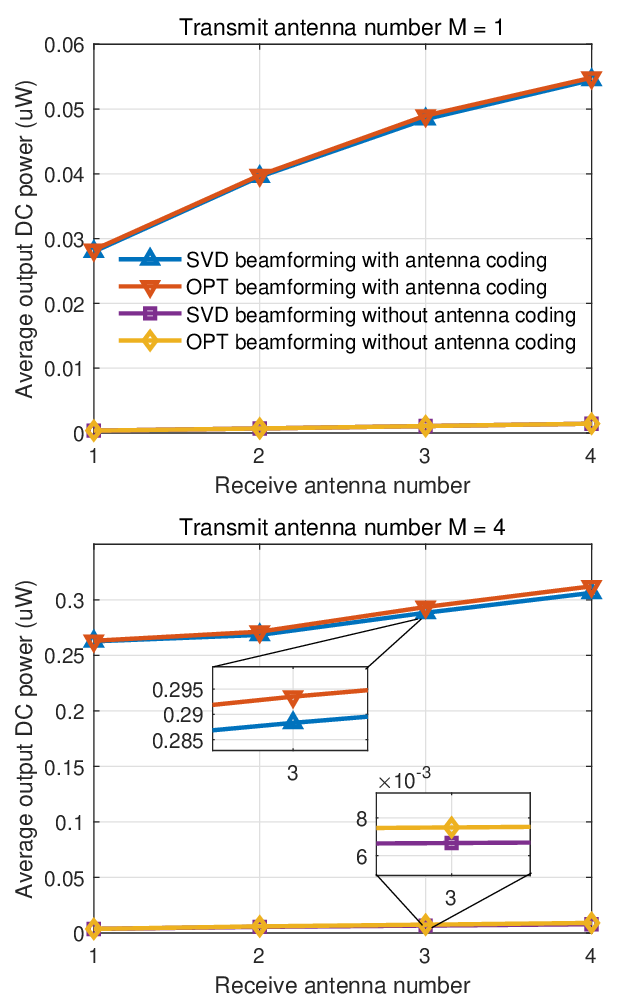} 
    \caption{Average output DC power versus receive antenna number for SIMO (top) and MIMO (bottom) configurations.}
    \label{result} 
\end{figure}

In Fig.~\ref{result}, the average output DC power is plotted versus the receive antenna number for SIMO and MIMO scenarios. The following observations can be made:

\textit{First}, using multiple transmit/receive pixel or conventional antennas is beneficial for increasing the output DC power of the WPT system with different beamforming strategies. With increasing number of receive antennas, the pixel antenna system performance increases with a steeper slope and more pronounced gain compared with the conventional system with fixed antenna configuration. 

\textit{Second}, regardless of the number of transmit/receive antennas, the MIMO WPT system with pixel antennas greatly outperforms the conventional system. As the number of transmit/receive antennas increases, the performance gap between pixel antennas and fixed antennas widens, indicating that pixel antennas benefit more significantly from additional antennas. This enhancement is due to the highly-reconfigurable radiation patterns adaptive to the multi-antenna channel matrix, so that the multiple paths between the equivalent spatially separated antennas can be coherently added with optimized pattern coders. The pixel antenna system can achieve up to 15 dB gain in output DC power for OPT beamforming with a $4\times4$ MIMO configuration, highlighting the significance of additional degrees of freedom introduced by the pixel antennas. Meanwhile, the pixel antenna system does not require additional RF chains or rectifier circuits, which yields low hardware complexity and power loss for scaling up the system. 

\textit{Third}, for both pixel and conventional antennas, the OPT beamforming has higher output DC power than SVD beamforming with MIMO configuration. The latter is suboptimal for output DC power without consideration of nonlinearities. As the RF-to-DC conversion efficiency increases with the input RF power, the rectenna nonlinearity should be properly leveraged in WPT to enhance the output DC power.

\section{Conclusion}
 In this work, we study the MIMO WPT system with antenna coding based on pixel antennas to boost the output DC power level. First, we propose the antenna coding model and beamspace channel model to demonstrate antenna coding as a new degree of freedom. Then, we formulate the DC power maximization problem with DC combining scheme and propose the alternating optimization algorithm for antenna coding and transmit beamforming design. Finally, we evaluate the proposed schemes through extensive simulations. By jointly exploiting gains from antenna coding, transmit beamforming, and rectenna nonlinearity, significant enhancement (15 dB) in output DC power compared with conventional systems is achieved, verifying the benefits of exploring WPT system design with pixel antennas. 

\section*{Acknowledgment}

The authors would like to acknowledge the support of the Hong Kong Research Grants Council for the General Research Fund (GRF) grant, 16208124, and the Science and
Technology Development Fund, Macau SAR (File/Project no. 001/2024/SKL and CG2025/IOTSC).


\end{document}